\documentclass[preprint,showpacs,preprintnumbers,amsmath,amssymb,aps]{revtex4}

\usepackage{graphicx}% Include figure files
\usepackage{dcolumn}% Align table columns on decimal point
\usepackage{bm}% bold math

\begin{document}

\title{Global quark polarization in non-central $A+A$ collisions}
\author{Jian-Hua Gao$^{1}$, Shou-Wan Chen$^{2}$,
Wei-tian Deng$^{1,3}$, Zuo-Tang Liang$^{1}$, Qun Wang$^2$
and Xin-Nian Wang$^{3,1}$}
\affiliation{
$^1$Department of Physics, Shandong University, Jinan, Shandong 250100, China\\
$^2$Department of Modern Physics,
University of Science and Technology of China, Hefei, Anhui 230026, China\\
$^3$Nuclear Science Division, MS 70R0319,
Lawrence Berkeley National Laboratory, Berkeley, California 94720, USA}

\date{\today}

\preprint{LBNL-63515}

%\vspace{-1.5in}
%\vspace{1.4in}

\begin{abstract}
Partons produced in the early stage of non-central heavy-ion collisions
can develop a longitudinal fluid shear because of unequal local
number densities of participant target and projectile nucleons. Under
such fluid shear, local parton pairs with non-vanishing impact parameter
have finite local relative orbital angular momentum along the direction
opposite to the reaction plane. Such finite relative orbital
angular momentum among locally interacting quark pairs can lead to global
quark polarization along the same direction due to spin-orbital coupling.
Local longitudinal fluid shear is estimated within both Landau
fireball and Bjorken scaling model of initial parton production.
Quark polarization through quark-quark scatterings with the exchange
of a thermal gluon is calculated beyond small-angle scattering
approximation in a quark-gluon plasma. The polarization is shown to have
a non-monotonic dependence on the local relative orbital angular momentum
dictated by the interplay between electric and magnetic interaction. It
peaks at a value of relative orbital angular momentum which scales
with the magnetic mass of the exchanged gluons. With the estimated
small longitudinal fluid shear in semi-peripheral $Au+Au$ collisions
at the RHIC energy, the final quark polarization is found to be
small $|P_q|<0.04$ in the weak coupling limit. Possible behavior of
the quark polarization in the strong coupling limit and implications
on the experimental detection of such global quark polarization
at RHIC and LHC are also discussed.
\end{abstract}


\maketitle

%\begin{widetext}

%\begin{multicols}{2}

\section{Introduction}

%Polarization effects in high energy reactions usually provide
%useful information on the reaction mechanism and often give us surprises.
%Such effects have been studied extensively in high energy
%lepton-hadron, hadron-hadron, and hadron-nucleus collisions
%and lead to an active field of High Energy Spin Physics.
%In contrast, less study has been
%made in this direction in high energy heavy-ion collisions.
%One of the reasons might be that it would be very difficult
%or even impossible to polarize a heavy ion beam.

Collective phenomena and jet quenching as observed in high-energy
heavy-ion collisions at the Relativistic Heavy-ion Collider (RHIC)
at the Brookhaven National Laboratory (BNL) provide
strong evidence for the formation of a strongly coupled quark
gluon plasma \cite{Gyulassy:2004zy,Jacobs:2004qv}.
Elliptic flow or azimuthal anisotropy of the hadron spectra in
semi-peripheral heavy-ion collisions and its agreement with
the ideal hydrodynamic calculations \cite{Ackermann:2000tr}
indicate a near perfect fluid behavior of the produced dense
matter. Such an empirical observation of small shear
viscosity \cite{Teaney:2003kp} is consistent with the large value of
the jet transport parameter as extracted from jet quenching study
of both single and dihadron spectra suppression \cite{Majumder:2007zh}.
Study of the collective behavior is made possible by investigating
hadron spectra in central rapidity region in non-central
or semi-peripheral heavy-ion collisions.
Extending the study to large rapidity region of non-central
heavy-ion collisions should provide more information not only
about the initial condition for the formation of the dense
matter \cite{Adil:2005qn} but also the dynamical properties of
the strongly coupled quark-gluon plasma.

Considering the longitudinal momentum distribution at various
transverse positions in a non-central heavy-ion collision, one
will find a longitudinal fluid shear distribution representing
local relative orbital angular momentum. Recently, it has been
pointed out that the presence of such local orbital angular
momentum of the partonic system at
the early stage of non-central heavy-ion collisions can lead to
a global polarization of quarks and anti-quarks
\cite{Liang:2004ph} in the direction orthogonal to the reaction plane.
Understanding the spin-orbital
interaction inside a strongly coupled system can open a new
window to the properties of quark-gluon-plasma (QGP).
Although no detailed calculations have been
carried out, an estimate using a screened static potential
model in the small angle approximation shows qualitatively that
spin-orbital coupling in Quantum Chromodynamics (QCD) can
lead to a finite global quark and anti-quark polarization.
Such a global quark/anti-quark polarization
should have many observable consequences such as global hyperon
polarization \cite{Liang:2004ph,Betz:2007kg}
and vector meson spin alignment \cite{Liang:2004xn}.
Predictions have been made \cite{Liang:2004ph,Liang:2004xn}
for these measurable quantities as functions of the global
quark polarization $P_q$ in various hadronization scenarios.
Since the reaction plane in heavy-ion
collisions can be determined in experiments by measuring the
elliptic and direct flows, measurements of the global
hyperon polarization or vector meson spin alignment
become feasible. These measurements at RHIC
are being carried out and some of the preliminary
results have already been reported
\cite{STARpol1,STARpol2,STARpol3,STARpol4,STARpol5,STARpol6,Abelev:2007zk}.

%The main idea and results in Ref. \cite{Liang:2004ph}
%can be summarized as follows.
%In a non-central high energy $AA$ collision,
%the system of nucleons participating the interaction
%in the overlapping region carries
%a large global orbital angular momentum along the
%direction orthogonal to that of the reaction plane.
%If a parton system such as a QGP is produced after the collision,
%the orbital angular momentum will be distributed among the partons,
%and the neighbouring partons in the system should have
%on average a local orbital angular momentum in the same direction.
%If the partons interact with each other,
%this local orbital angular momentum can be transferred
%to the polarization of quarks and anti-quarks due to the
%spin-orbital coupling in QCD,
%leading to a globally polarized quark-gluon-plasma.

The estimate of the global quark polarization in Ref. \cite{Liang:2004ph}
was obtained by evaluating the polarization cross section
in the impact parameter space with small angle approximation
in an effective potential model. The analytical result,
\begin{equation}
P_q=-\pi\mu p/2E(E+m_q)
\end{equation}
has an intuitive expression, where $p$ is the average {\em c.m.} momentum
of two partons with an average transverse separation $1/\mu$ due to
the longitudinal fluid shear. However, for a massless quark in a
small longitudinal fluid shear, the obtained
quark polarization $P_q$ can become larger than 1, indicating
the breakdown of the small angle approximation.
A more realistic estimate in non-central
heavy-ion collisions at RHIC indicates a small value of the
average longitudinal fluid shear. Therefore, it is imperative to have a
more realistic estimate of the quark polarization $P_q$ beyond
the small angle approximation. This will be the focus of this
paper.

The rest of the paper is organized as follows. In Sec.~II, we
calculate the average longitudinal fluid shear in two different
models of parton production. In a Landau fireball picture, a wounded nucleon
model of bulk parton production in heavy-ion collisions is used
with both simple hard-sphere and more realistic Wood-Saxon nuclear geometry.
In the Bjorken scaling scenario, we use HIJING Monte Carlo
model to estimate the transverse shear of the rapidity distribution
of the produced parton in heavy-ion collisions at the RHIC energy
which will be used to estimate the longitudinal fluid shear in the
local comoving frame of the plasma.
In Sec. III, we use the Hard Thermal Loop (HTL) resummed gluon propagator
in the comoving frame of the local longitudinal fluid cell to
extend the calculation of quark polarization in Ref. \cite{Liang:2004ph}
beyond small angle approximation and discuss the relative contributions from
electric and magnetic part of quark-quark scattering. Finally in
Sec. IV, we discuss the numerical results and their implications
for experimental measurements at RHIC.

\section{Orbital angular momentum and shear flow}

\begin{figure}[htbp]
\includegraphics[width=6cm]{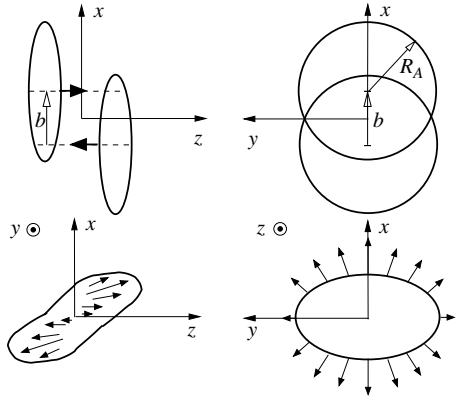}
%\centerline{\psfig{figure=geo.eps,width=2.5in,height=3.1in}}
\caption{Illustration of a non-central heavy-ion collision with
impact parameter $\vec{b}$.
The global angular momentum of the produced dense matter
is along $-\hat{y}$, opposite to the reaction plane.}
\label{fig1}
\end{figure}

Let us consider two colliding nuclei with the projectile of
beam momentum per nucleon $\vec p_{in}$ moving in the direction of the $z$
axis, as illustrated in Fig.~\ref{fig1}. The impact parameter $\vec{b}$,
defined as the transverse distance of the center of the projectile nucleus
from that of the target, is taken to be along
the $\hat{x}$-direction. The normal direction $\vec{n}_b$ of the
reaction plane, given by,
\begin{equation}
\vec{n}_b\equiv \vec p_{in} \times \vec{b}/|\vec p_{in}\times\vec{b}|,
\label{eq:rplane}
\end{equation}
is along $\hat{y}$. For a non-central collision, the two colliding nuclei
carry a finite global orbital angular momentum $L_y$
along the direction orthogonal to the reaction plane ($-\hat{y}$).
How such a global orbital angular momentum is transferred to the
final state particles depends on the equation of state (EOS)
of the dense matter. At low energies, the final state is expected
to be the normal nuclear matter with an EOS of rigid nuclei.
A rotating compound nucleus can be formed when the colliding energy
is comparable or smaller than the nuclear binding energy. The
finite value of the total orbital angular
momentum of the non-central collision at such low energies
provides a useful tool for the study of the properties of
superdeformed nuclei under such rotation \cite{Cederwall:1994gz}.
At high colliding energy at RHIC, the dense matter is expected to be
partonic with an EOS of the quark-gluon plasma. Given such a soft
EOS, the global orbital angular momentum would probably
never lead to the global rotation of the dense matter. Instead,
the total angular momentum will be distributed across the overlapped
region of nuclear scattering and is manifested in the shear of
the longitudinal flow leading to a finite value of local
vorticity density. Under such longitudinal fluid shear, a pair
of scattering partons will on average carry a finite value
of relative orbital angular momentum in the opposite direction to the
reaction plane as defined in Eq. (\ref{eq:rplane}). According to
Ref. \cite{Liang:2004ph}, quark (or antiquark) will acquire
a global polarization after such scatterings through the spin-orbital
coupling in QCD.

The magnitude of the total orbital angular momentum $L_y$ and the
resulting longitudinal fluid shear can both be estimated within
the wounded nucleon model of particle production
in which the number of produced particles is assumed to
be proportional to the number of participant nucleons. The
transverse distributions (integrated over $y$) of participant
nucleons in each nucleus can be written as,
\begin{equation}
\label{dndx}
\frac{dN_{\rm part}^{P,T}}{dx}
=\int dydz \rho_A^{P,T}(x,y,z,b),
\end{equation}
in terms of the participant nucleon number density  $\rho_A^{P,T}(x,y,z,b)$
in nucleus $A$ in the coordinate system defined above.
The superscript $P$ or $T$ denotes projectile or target respectively.
The total orbital angular momentum $L_y$ of the two colliding nuclei
can be defined as,
\begin{equation}
\label{ly}
L_y=-p_{in}\int x\ dx (\frac{dN_{\rm{part}}^P}{dx}
-\frac{dN_{\rm{part}}^T}{dx}).
\end{equation}
where $p_{in}$ is the momentum per incident nucleon.

We assume the Woods-Saxon nuclear distribution,
\begin{equation}
f_{WS}^{P,T}(x,y,z,b)=C\left( 1
+\exp {\frac{\sqrt{(x\mp b/2)^2+y^2+z^2}-R_A}{a}}\right)^{-1} \,.
\end{equation}
The participant nucleon number density is then
\begin{equation}
\rho_{A,WS}^{P,T}(x,y,z,b)=f_{WS}^{P,T}(x,y,z,b)
\left\{ 1-\exp
\left[ -\sigma_{NN}\int dz f_{WS}^{T,P}(x,y,z,b)\right]\right\} \, ,
\end{equation}
where $\sigma_{NN}\approx 42$  mb is the total cross section of
nucleon-nucleon scatterings at the RHIC energy, $C$ is the
normalization constant and $a$ is the width parameter set to $a=0.54$ fm.

\begin{figure}[htbp]
%\resizebox{3.8in}{2.6in}
\includegraphics[width=11cm]{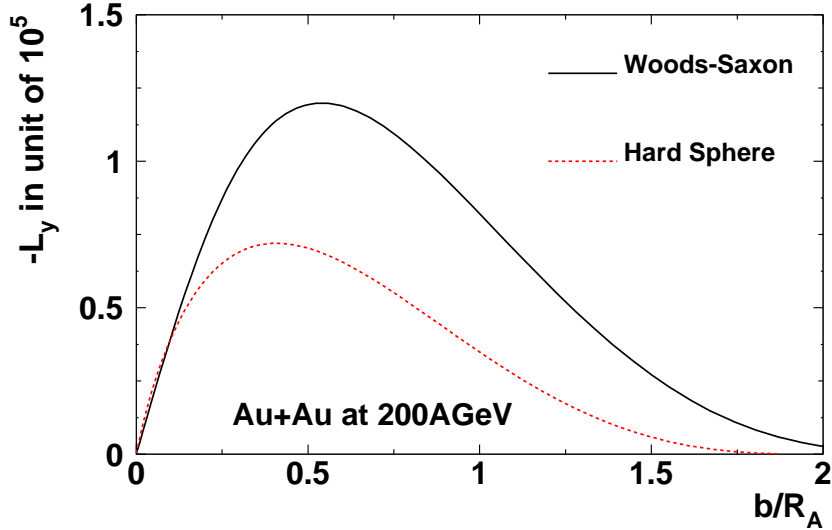}
\caption{(Color online) The total orbital angular momentum of the overlapping
system in $Au+Au$ collisions at the RHIC energy
as a function of the impact parameter $b$.
The solid and dashed curves are from the Woods-Saxon and hard-sphere
distributions, respectively.}
\label{fig2}
\end{figure}

Shown in Fig.~\ref{fig2} as the solid line is the numerical value
of $L_y$ as a function of $b$ for the Woods-Saxon nuclear
distribution. As a comparison, we also plot as the dashed line
the $L_y$ distribution with a hard-sphere nuclear
distribution which was used in Ref. \cite{Liang:2004ph}.
With the hard-sphere nuclear distribution, the participant
nucleon density is given by the overlapping area of two
hard spheres,
\begin{equation}
\label{hard-sphere1}
\rho _{A,HS}^{P,T}(x,y,z,b)
=f_{A,HS}^{P,T}(x,y,z,b)\theta(R_A-\sqrt{(x\pm b/2)^2+y^2+z^2}),
\end{equation}
and
\begin{equation}
\label{hard-sphere}
f_{A,HS}^{P,T}(x,y,z,b)
=\frac{3A}{4\pi R_A^3}\theta(R_A-\sqrt{(x\mp b/2)^2+y^2+z^2}),
\end{equation}
where $R_A=1.12A^{1/3}$ is the nuclear radius and $A$ the atomic number.
We note that there are significant differences between two nuclear
geometry in the total orbital angular momentum $L_y$ in the overlapped
region of two colliding nuclei. In both cases, the total orbital angular
momentum is huge and is of the order of $10^4$ at most impact parameters.

\begin{figure}[ht]
\includegraphics[width=10cm]{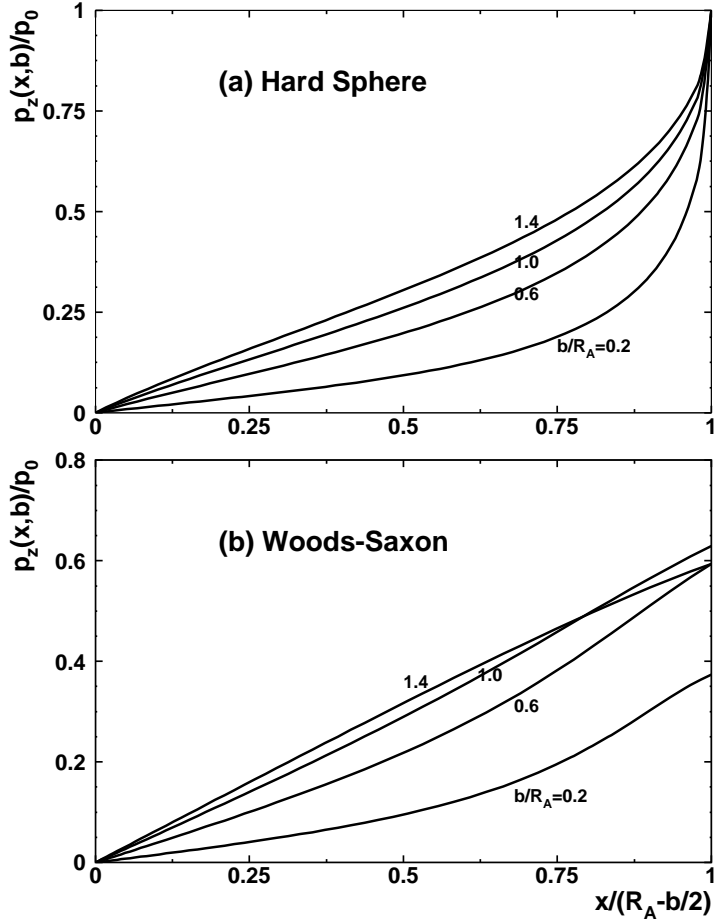}
\caption{The average longitudinal momentum distribution $p_z(x,b)$ in
unit of $p_0=\sqrt{s}/[2c(s)]$ as a function of $x/(R_A-b/2)$ for
different values of $b/R_A$ with the hard sphere (upper panel)
and Woods-Saxon (lower panel) nuclear distributions.}
\label{fig3}
\end{figure}

Since RHIC data indicate the formation of a
strongly coupled quark-gluon plasma \cite{Gyulassy:2004zy},
we can assume that a partonic system is formed immediately following
the initial collision and interactions among partons will
lead to both transverse (in $x$-$y$ plane) and longitudinal collective motion
in the quark-gluon plasma (QGP). The total orbital angular momentum
carried by the produced system will manifest in
the longitudinal flow shear or a finite value of the
transverse (along $\hat{x}$) gradient of the longitudinal flow velocity.
How the total angular momentum is distributed to the longitudinal
flow shear and the magnitude of the local relative orbital angular
momentum depends on the parton production mechanism and their
longitudinal momentum distributions. We consider two different scenarios
in this paper: Landau fireball and Bjorken scaling model.

By momentum conservation, the average initial collective longitudinal
momentum at any given transverse position can be calculated
as the total momentum difference between participating projectile
and target nucleons. Since the total multiplicity in $A+A$ collisions
is proportional to the number of participant nucleons \cite{phobos2},
we can make the same assumption for the produced partons
with a proportionality constant $c(s)$ at a given
center of mass energy $\sqrt{s}$.

In a Landau fireball model, we assume the produced partons thermalize
quickly and have a common longitudinal flow velocity at a given transverse
position of the overlapped region. The average collective longitudinal
momentum per parton can be written as
\begin{equation}
p_z(x,b;\sqrt{s})=p_0\frac{dN_{\rm part}^P/dx - dN_{\rm part}^T/dx}
{dN_{\rm part}^P/dx + dN_{\rm part}^T/dx},
\end{equation}
where $p_0=\sqrt{s}/[2c(s)]$. The distribution $p_z(x,b;\sqrt{s})$ is an odd
function in both $x$ and $b$ and therefore vanishes
at $x=0$ or $b=0$. In Fig.~\ref{fig3},
$p_z(x,b;\sqrt{s})$ is plotted as a function
of $x$ at different impact parameters $b$. We see that
$p_z(x,b;\sqrt{s})$ is a monotonically increasing function of
$x$ until the edge of the overlapped region
$|x\pm b/2|=R_A$ beyond which it drops to zero (gradually for Woods-Saxon
geometry).

From $p_z(x,b;\sqrt{s})$ one can compute the
transverse gradient of the average longitudinal collective
momentum per parton $dp_z/dx$ which is an even function
of $x$ and vanishes at $b=0$. One can then estimate the longitudinal
momentum difference $\Delta p_z$ between two neighboring partons
in QGP. On average, the relative orbital angular momentum
for two colliding partons separated by $\Delta x$ in the transverse direction
is $l_y \equiv -(\Delta x)^2dp_z/dx$. With the hard sphere nuclear
distribution, $l_y$ is proportional
to $dp_0/dx\equiv p_0/R_A=\sqrt{s}/[2c(s)R_A]$.

\begin{figure}[ht]
\includegraphics[width=10cm]{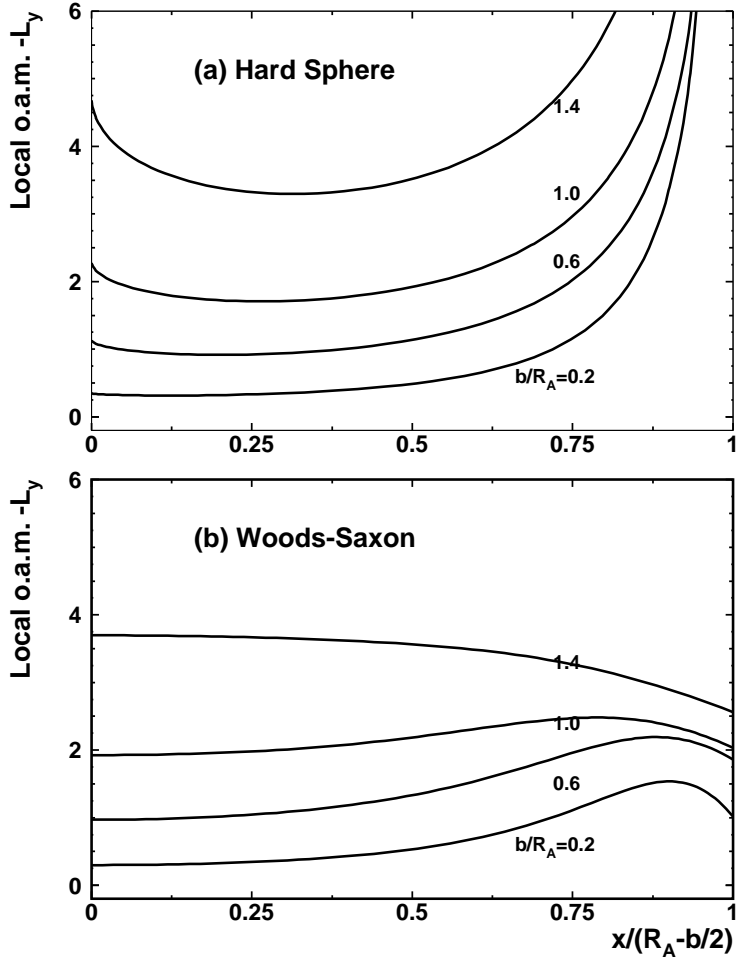}
\caption{The average orbital angular momentum
$l_y \equiv -(\Delta x)^2dp_z/dx$
of two neighboring partons separated by $\Delta x=1$ fm as a
function of the scaled transverse coordinate $x/(R_A-b/2)$ for
different values of the impact parameter $b/R_A$ with the hard-sphere (upper panel)
and Woods-Saxon (lower panel) nuclear distributions.}
\label{fig:ly}
\end{figure}

In $Au+Au$ collisions at $\sqrt{s}=200$ GeV, the number of charged
hadrons per participating nucleon is about 15 \cite{phobos2}.
Assuming the number of partons per (meson dominated)
hadron is about 2, we have $c(s)\simeq 45$ (including neutral hadrons).
Given $R_A=6.5$ fm, $dp_0/dx \simeq 0.34$ GeV/fm
and we obtain $l_0\equiv -(\Delta x)^2 dp_0/dx \simeq -1.7$
for $\Delta x=1$ fm.
In Fig. \ref{fig:ly}, we show the average local orbital angular momentum
$l_y$ for two neighboring partons separated by $\Delta x=1$ fm
as a function of $x$ for different impact parameter $b$ for both
Woods-Saxon and hard-sphere nuclear distributions.
We see that $l_y$ is in general of the order of 1 and
is comparable or larger than the spin of a quark.
%So the effect can indeed be very significant.
%We also note that $l_y$ does not depend on $A$ but it
%depends on $\sqrt{s}$.
It is expected that $c(s)$ should depend logarithmically
on the colliding energy $\sqrt{s}$,
therefore $l_y$ should increases with growing $\sqrt{s}$.

In a 3-dimensional expanding
system, there could be strong correlation
between longitudinal flow velocity and spatial coordinate of the
fluid cell. The most simplified picture is the Bjorken scaling
scenario \cite{Bjorken:1982qr} in which the longitudinal flow
velocity is identical to the spatial
velocity $\eta=\log[(t+z)/(t-z)]$. With such correlation,
local interaction and thermalization requires that a parton
only interacts with other partons in the same region of longitudinal
momentum or rapidity $y$. The width of such region in rapidity is
determined by the half-width of the thermal distribution
$f(Y,p_T)=\exp[-p_T\cosh(Y-\eta)/T]$ \cite{Levai:1994dx}, which is
approximately $\Delta_Y\approx 1.5$ (with $\langle p_T\rangle\approx 2T$).
The relevant measure of the local relative orbital angular momentum
between two interacting partons is, therefore, the difference in parton
rapidity distributions at transverse distance of $\delta x\sim 1/\mu$ on the
order of the average interaction range.

\begin{figure}[ht]
\includegraphics[width=10cm]{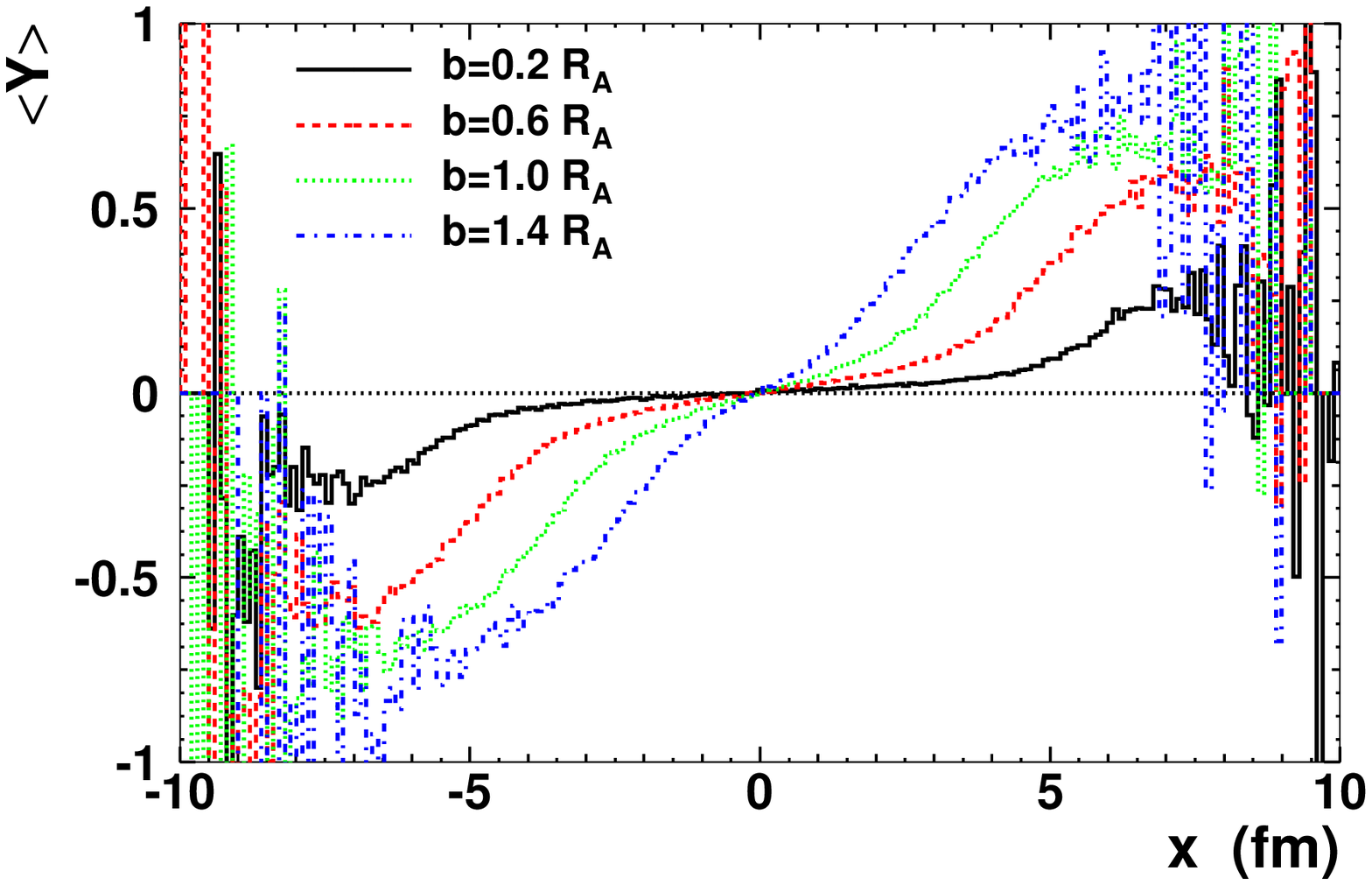}
\caption{(Color online) The average rapidity $\langle Y\rangle$ of the
final state particles as a function of the transverse coordinate $x$ from HIJING
Mont Carlo simulations \cite{Wang:1991ht,Wang:1996yf} of non-central
$Au+Au$ collisions at $\sqrt{s}=200$ GeV.}
\label{yvsx}
\end{figure}

\begin{figure}[ht]
\includegraphics[width=10cm]{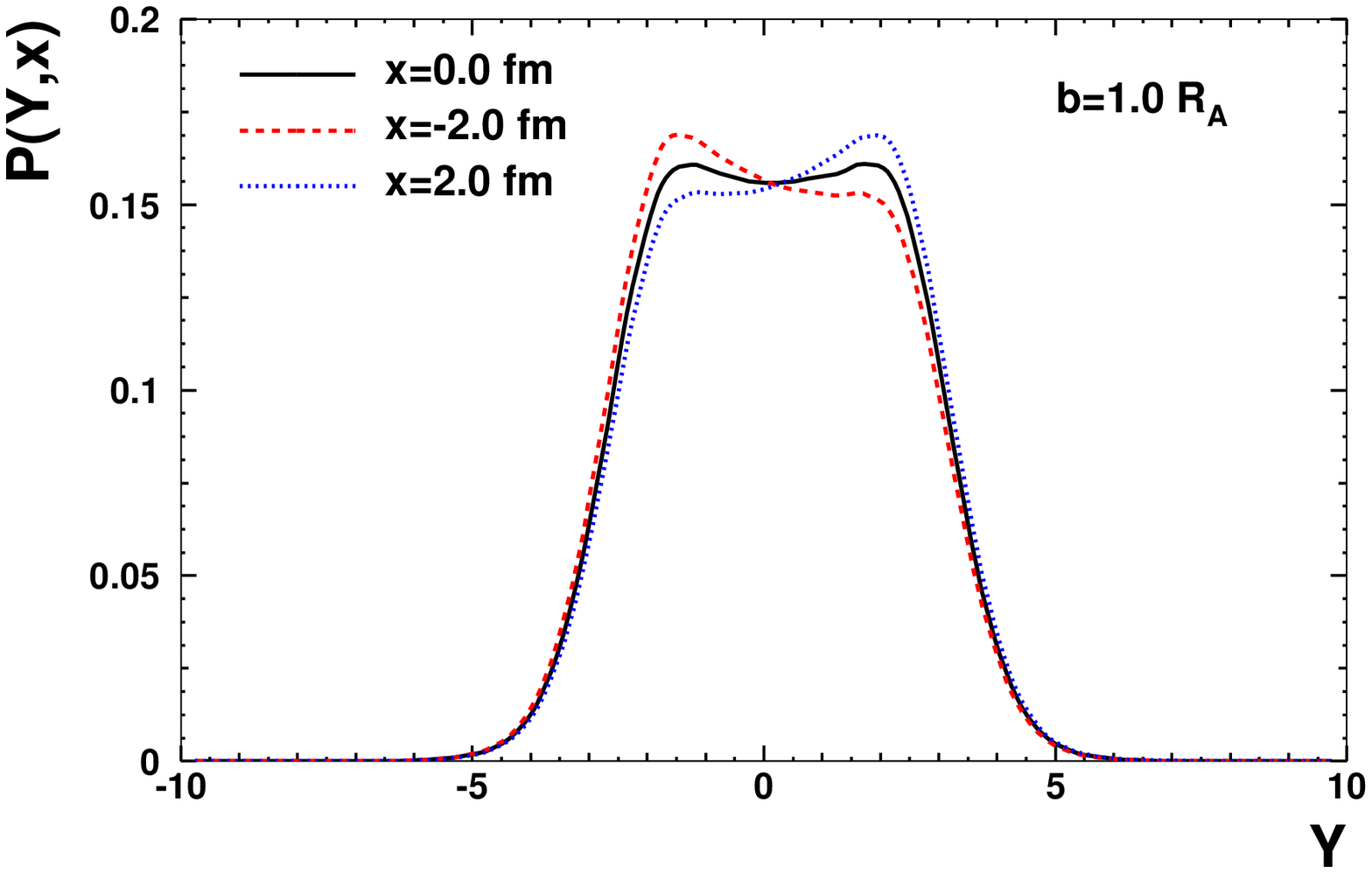}
\caption{(Color online) The normalized rapidity distribution of
particles $P(Y,x)$ [Eq. (\ref{eq:pxy})] at different transverse
position $x$ from HIJING simulations of non-central $Au+Au$ collisions
at $\sqrt{s}=200$ GeV.}
\label{dndy}
\end{figure}

One needs a dynamical model to estimate the local rapidity distributions
of produced partons. For such a purpose, we use
HIJING Monte Carlo model \cite{Wang:1991ht,Wang:1996yf} to
calculate the hadron rapidity distributions at different transverse
coordinate ($x$) and assume that parton distributions of the dense matter
are proportional to the final hadron spectra. Shown in Fig.~\ref{yvsx}
is the average rapidity $\langle Y\rangle $ as
a function of the transverse coordinate $x$ for different values of
the impact parameter $b$. The distributions have exactly the same
features as given by the wounded nucleon model in Fig.~\ref{fig3}.
The variation of the rapidity distributions with respect to the
transverse coordinate is illustrated in Fig.~\ref{dndy} by the
normalized rapidity distributions
\begin{equation}
P(Y,x)=\frac{dN/dxdY}{dN/dx}
\label{eq:pxy},
\end{equation}
at different transverse coordinates, $x=0,\pm 2$ fm. At finite values
of the transverse coordinates $x$, the normalized rapidity distributions
evidently peak at larger values of rapidity $|Y|$. The shift in
the shape of the rapidity distributions will provide the local longitudinal
fluid shear or finite relative orbital angular momentum for
two interacting partons in the local comoving frame at any given
rapidity $Y$. To quantify such longitudinal fluid shear, one
can calculate the average rapidity within an interval $\Delta_Y$
at $Y$,
\begin{equation}
\langle Y\rangle \approx Y +\frac{\Delta_Y^2}{12}\frac{1}{P(Y,x)}
\frac{\partial P(Y,x)}{\partial Y} .
\end{equation}
The average rapidity shear or the difference in average rapidity for
two partons separated  by a unit of transverse distance $\Delta x=1$ fm
is then,
\begin{equation}
\frac{\partial \langle Y\rangle}{\partial x} \approx
\frac{\Delta_Y^2}{12} \frac{\partial^2 \log P(Y,x)}{\partial Y\partial x}.
\end{equation}

%the fluid shear
%in the local comoving frame at given rapidity $y$ is finite
%and it peaks at large value of rapidity $|y|\approx 2$. It is also
%generally smaller than the averaged fluid shear in the center of mass
%frame of two colliding nuclei in the Landau fireball model.

\begin{figure}[ht]
\includegraphics[width=10cm]{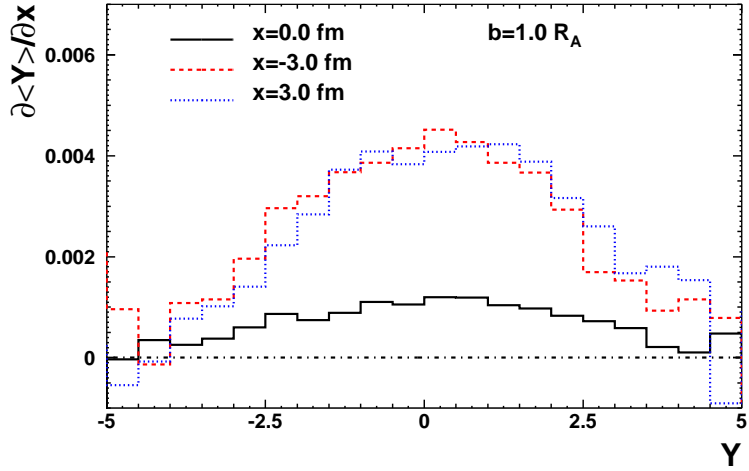}
\caption{(Color online) The average rapidity shear
$\partial \langle Y\rangle/\partial x$ within
a window $\Delta_Y=1$ as a function of the rapidity $Y$ at
different transverse position $x$ from HIJING calculation of
non-central $Au+Au$ collisions at $\sqrt{s}=200$ GeV.}
\label{dndydx}
\end{figure}

Shown in Fig.~\ref{dndydx} is the average rapidity shear as a function
of the rapidity $y$ at different values of the transverse coordinate $x$
for $\Delta_Y=1$.
As we can see, the average rapidity shear has a positive and finite
value in the central rapidity region. The corresponding local
relative longitudinal momentum shear is
\begin{equation}
\frac{\partial \langle p_z\rangle}{\partial x}\approx p_T\cosh Y
\frac{\partial \langle Y\rangle}{\partial x} \,\, .
\end{equation}
With $\langle p_T\rangle \approx 2T\sim 0.8$ GeV, we have
%the above average longitudinal momentum shear
$\partial \langle p_z\rangle/\partial x\sim 0.003$
GeV/fm in the central rapidity region of a non-central $Au+Au$ collision
at the RHIC energy given by the HIJING simulations,
which is much smaller than that from a Landau fireball model estimate.

%\begin{figure}[ht]
%\includegraphics[width=10cm]{dpzdx-hs.eps}
%\includegraphics[width=10cm]{dpzdx-ws.eps}
%\caption{The gradient $dp_z/dx$ in unit of $2p_0/R_A$
%as a function of $x/(R_A-b/2)$ for
%different values of $b/R_A$.}
%\label{fig4}
%\end{figure}

\section{Global quark polarization}

As we have discussed earlier, under the longitudinal fluid shear,
a pair of interacting parton will have a finite value of relative
orbital angular momentum along the direction opposite to the
reaction plane. In this section, we will calculate quark
polarization via scatterings with fixed direction of the relative
orbital angular momentum. We will assign a fixed direction of
the impact parameter $\vec x_T$ between two interacting partons
to reflect the direction of the
relative orbital angular momentum. The magnitude of the relative
orbital angular momentum will be charaterized by the relative
longitudinal momentum $p$ between two partons separated
by a transverse distance $\Delta x\sim 1/\mu$ on the order of
the average interaction range. With the averaged longitudinal fluid
shear $dp_z/dx$ in the center of mass frame of the two colliding
nuclei in the Landau fireball model, we have $p=\Delta x (dp_z/dx)$.
In the Bjorken scaling scenario with
strong correlation between spatial and momentum rapidity,
the average local longitudinal shear in the comoving frame
will be given by
$p=\Delta x p_T\cosh(Y) \partial \langle Y \rangle/\partial x$,
where $p_T$ is the average transverse momentum.

%in Sec. II, in a non-central high energy $AA$ collision,
%there exists a huge global orbital angular momentum for
%the partonic system. The global orbital angular momentum implies the existence
%of a local orbital angular momentum $l_y$ for two neighboring
%partons in the system. If the partons interact with each other,
%the local relative orbital angular momentum $l_y$ can
%be converted to quark polarization via spin-orbital coupling.
%In other words, quarks can be polarized after scatterings
%with fixed direction of relative orbital angular momentum.
%We know that spin-orbital coupling is an intrinsic property of any
%relativistic theory. It is thus interesting
%to make such a calculation to see how large
%the polarization can be.
%The physical picture where we start is as follows.
%The QGP is supposed to be formed after $AA$ collisions.
%Because the collision is usually non-central, the QGP has
%a global orbital angular momentum, thus two neighboring partons
%have on average a local orbital angular momentum $l_y$ in the $-y$ direction.
%In medium partonic interaction can be described by QCD at finite temperature.
%We are aimed to compute the polarization in a single parton-parton scattering.

%Since fixed direction of relative orbital angular momentum implies
%a preferred direction of impact parameter between the two partons,
%we consider quark scatterings with fixed impact parameter $\vec{x}_T$.
%This can be carried out in the following way.

\subsection{Quark scattering at fixed impact parameter}

We consider the scattering
$q_1(P_1,\lambda_1)+q_2(P_2,\lambda_2)\to q_1(P_3,\lambda_3)+q_2(P_4,\lambda_4)$
of two quarks with different flavors,
where $P_i=(E_i,\vec p_i)$ and $\lambda_i$ in the brackets denote
the four momenta and spins of the quarks respectively.
The cross section in momentum space is given by,
\begin{equation}
d\sigma_{\lambda_3}=\frac{c_{qq}}{F}\frac{1}{4}\sum_{\lambda_1,\lambda_2,\lambda_4}
\mathcal{M}(Q){\mathcal{M}}^{*}(Q)
(2\pi)^{4}\delta(P_1+P_2-P_3-P_4)\frac{d^3{\vec{p}}_3}{(2\pi)^{3}2E_3}
\frac{d^3{\vec{p}}_4}{(2\pi)^{3}2E_4},\\
\end{equation}
where ${\cal{M}}(Q)$ is the scattering amplitude in momentum space,
$Q=P_3-P_1=P_2-P_4$ is the four momentum transfer,
$c_{qq}=2/9$ is the color factor, and
$F=4\sqrt{(P_1\cdot P_2)^2-m_1^2m_2^2}$ is the flux factor.
Since we are interested in the polarization of one of the quarks
$q_1$ after the scattering,
we therefore average over the spins
of initial quarks and sum over
the spin of the quark $q_2$ in the final state.

We work in the center of mass frame of the two quark system. For
simplification, we neglect thermal momentum in the transverse direction
and assume the relative
momentum of the two quarks separated by a transverse distance
$\Delta x$ of the order of the effective interaction range $1/\mu$
is simply given by the longitudinal fluid shear.

One can integrate over the final momentum $\vec{p}_4$ of the quark $q_2$
and the longitudinal component $p_{3z}$ of the quark $q_1$,
and obtain
\begin{equation}
d\sigma_{\lambda_3}=\frac{c_{qq}}{4F}\frac{1}{4}\sum_{\lambda_1,\lambda_2,\lambda_4}
\sum_{i=+,-}
\frac{1}{E_{2}|p_{3z}^i| +E_{1}|p_{3z}^i|}
%\mathcal{M}({\vec{q}}_{T}){\mathcal{M}}^{*}({\vec{q}}_{T})
\mathcal{M}(Q_i){\mathcal{M}}^{*}(Q_i)
\frac{d^2{\vec{q}}_{T}}{(2\pi)^{2}},
\end{equation}
where $p_{3z}^{\pm}=\pm \sqrt{p^2-{q_{T}}^2}$, corresponding to two possible
solutions of the energy-momentum conservation in the
elastic scattering process,
$p=|\vec{p}_1|=|\vec{p}_2|$;
and  $\vec q_T=\vec{p}_{3T}$ is the transverse momentum transfer.
For simplicity, we will suppress the summation notation over $i=+,-$
hereafter but keep in mind that the final cross section includes the two terms.

Since we would like to calculate the polarization of one final-state
quark with a fixed direction of the orbital angular momentum, or
fixed direction of the impact parameter,  we will cast the cross
section in impact parameter space by making a two dimensional Fourier
transformation in the transverse momentum transfer $\vec q_T$, {\em i.e.},
\begin{equation}
\frac{d^2\sigma_{\lambda_3}}{d^{2}{\vec{x}}_{T}}
=\frac{c_{qq}}{16F}\sum_{\lambda_1,\lambda_2,\lambda_4}
\int\frac{d^{2}{\vec{q}}_{T}}{(2\pi)^{2}}\frac{d^{2}{\vec{k}}_{T}}{(2\pi)^{2}}
e^{i({\vec{k}}_{T}-{\vec{q}}_{T})\cdot{\vec{x}}_{T}}
\frac{\mathcal{M}({\vec{q}}_{T})}{\Lambda({\vec{q}}_{T})}
\frac{{\mathcal{M}}^{*}({\vec{k}}_{T})}{{\Lambda}^{*}({\vec{k}}_{T})},
\end{equation}
where
${\mathcal{M}}({\vec{q}}_{T})$ and ${\mathcal{M}}({\vec{k}}_{T})$
are the scattering matrix elements in momentum space
with four momentum transfer $Q=(0,\vec{q})$ and $K=(0,\vec{k})$ respectively,
and
\begin{equation}
\Lambda({\vec{q}}_{T})=\sqrt{(E_1+E_2)|p_{3z}^+|}.
\end{equation}

To calculate the quark-quark scattering amplitude in a thermal
medium, we will use Hard Thermal Loop (HTL) resummed
gluon propagator \cite{WELD82,hw96},
\begin{equation}
\Delta^{\mu \nu}(Q)=\frac{P_T^{\mu\nu}}{-Q^2+\Pi_T(x)}
+\frac{P_L^{\mu\nu}}{-Q^2+\Pi_L(x)}+(\alpha-1)\frac{Q^{\mu}Q^{\nu}}{Q^4},
\label{eq:htl1}
\end{equation}
where $Q$ denotes the gluon four momentum and
$\alpha$ is the gauge fixing parameter.
The longitudinal and transverse projectors $P_{T,L}^{\mu\nu}$
are defined by
\begin{eqnarray}
  {P}_L^{\mu\nu}&=&\frac{-1}{Q^2q^2}(\omega Q^{\mu}-Q^2U^{\mu})
  (\omega Q^{\nu}-Q^2U^{\nu})\, , \\
  P_T^{\mu\nu}&=&\tilde{g}^{\mu\nu}
  +\frac{\tilde{Q}^{\mu}\tilde{Q}^{\nu}}{q^2}\, ,
\end{eqnarray}
with $\omega=Q\!\cdot\!U$, $\tilde{Q}_{\mu}=Q_{\mu}-U_{\mu}\omega$,
$q^2=-\tilde{Q}^2$, $\tilde{g}_{\mu\nu}=g_{\mu\nu}-U_{\mu}U_{\nu}$.
Here $U$ is the fluid velocity of the local medium.
The transverse and longitudinal self-energies are given by \cite{WELD82}
\begin{eqnarray}
\Pi_L(x)&=&\mu_D^2\left[1-\frac{x}{2}
\ln\left(\frac{1+x}{1-x}\right) + i \frac{\pi}{2}x \right](1-x^2)\,,
\label{eq:piL} \\
\Pi_T(x)&=&\mu_D^2\left[\frac{x^2}{2}+\frac{x}{4}(1-x^2)
\ln\left(\frac{1+x}{1-x}\right) - i \frac{\pi}{4}x(1-x^2)\right]
\label{eq:piT}
\end{eqnarray}
where $x=\omega/q$ and $\mu_D^2=g^2(N_c+N_f/2)T^2/3$ is the
Debye screening mass.

With the above HTL gluon propagator, the quark-quark scattering
amplitudes can be expressed as
\begin{equation}
\mathcal{M}({\vec{q}}_{T})= {\overline{u}}_{{\lambda}_3}(P_1+Q){\gamma}_{\mu}
u_{{\lambda}_1}(P_1) {\Delta}^{\mu\nu}(Q)
{\overline{u}}_{{\lambda}_4}(P_2-Q){\gamma}_{\nu}u_{{\lambda}_2}(P_2),
\end{equation}
\begin{equation}
{\mathcal{M}}^{*}({\vec{k}}_{T})=
{\overline{u}}_{{\lambda}_1}(P_1){\gamma}_{\alpha}
u_{{\lambda}_3}(P_1+K) {\Delta}^{\alpha\beta*}(K)
{\overline{u}}_{{\lambda}_2}(P_2){\gamma}_{\beta}
u_{{\lambda}_4}(P_2-K).
\end{equation}
The product ${\cal{M}}({\vec{q}}_{T}){\cal{M^*}}({\vec{k}}_{T})$ can be converted to the following
trace form,
\begin{eqnarray}
\label{trace}
\sum_{\lambda_1,\lambda_2}{\cal{M}}({\vec{q}}_{T}){\cal{M^*}}({\vec{k}}_{T})&=&
\Delta^{\mu\nu}(Q)\Delta^{\alpha\beta*}(K)
{\rm Tr}[u_{\lambda_3}(P_1+K)\bar u_{\lambda_3}(P_1+Q)
\gamma_\mu(P_1\hspace{-10pt}\slash+m_1)\gamma_\alpha]\nonumber\\
&&\times{\rm Tr}[u_{\lambda_4}(P_2-K)\bar u_{\lambda_4}(P_2-Q)
\gamma_\nu(P_2\hspace{-10pt}\slash+m_2)\gamma_\beta].
\end{eqnarray}

In calculations of transport coefficients such as jet energy loss
parameter \cite{screen} and thermalization time \cite{hw96} which
generally involve cross sections weighted with transverse momentum
transfer, the imaginary part of the HTL propagator in the magnetic
sector is enough to regularize the infrared behavior of the transport
cross sections. However, in our following calculation of quark polarization,
total parton scattering cross section is involved. The contribution
from the magnetic part of the interaction has therefore
infrared divergence which can only be regularized through the
introduction of non-perturbative magnetic screening mass
$\mu_m\approx0.255\sqrt{N_c/2} g^2T$~\cite{TBBM93}.

Since we have neglected the thermal momentum perpendicular to the
longitudinal flow, the energy transfer $\omega=0$ in the center of
mass frame of the two colliding partons. This corresponds to
setting $x=0$ in the HTL resummed gluon propagator in Eq. (\ref{eq:htl1}).
In this case, the center of mass frame of scattering quarks
coincides with the local comoving frame of QGP and
the fluid velocity is $U^{\mu}=(1,0,0,0)$. The corresponding
HTL effective gluon propagator in Feynman gauge that contributes to
the scattering amplitudes is reduced to,
\begin{equation}
{\Delta}^{\mu\nu}(Q)=
\frac{g^{\mu\nu}-U^\mu U^\nu}{q^2+\mu_m^2}
+\frac{U^\mu U^\nu}{q^2+{{\mu}_D}^2}.
\end{equation}

The differential cross section can in general be decomposed into a
spin-independent and a spin-dependent part,
\begin{equation}
\frac{d^2\sigma_{\lambda_3}}{d^{2}{\vec{x}}_{T}}=
\frac{d\sigma}{d^{2}{\vec x}_{T}}
+\lambda_3\frac{d\Delta\sigma}{d^{2}{\vec x}_{T}}.
\end{equation}
with
\begin{equation}
\frac{d^2\sigma}{d^{2}{\vec x}_{T}}
=\frac{1}{2}\left(\frac{d\sigma_+}{d^{2}{\vec x}_{T}}+\frac{d\sigma_-}{d^{2}{\vec x}_{T}}\right),
\end{equation}
\begin{equation}
\frac{d^2\Delta\sigma}{d^{2}{\vec x}_{T}}
=\frac{1}{2}\left(\frac{d\sigma_+}{d^{2}{\vec x}_{T}}-\frac{d\sigma_-}{d^{2}{\vec x}_{T}}\right).
\end{equation}

The spin-dependent part will mostly determine the polarization of the
final state quark $q_1$ via the scattering. The calculation is involved.
A simple estimate was given in Ref. \cite{Liang:2004ph},
using a screened static potential model and small angle approximation.
In this case, the cross sections can be written in a general form as,
\begin{eqnarray}
\label{csgform}
\frac{d^2\sigma}{d^{2}{\vec x}_{T}}&=&F(x_T,E), \\
\label{dcsgform}
\frac{d^2\Delta\sigma}{d^{2}{\vec x}_{T}}
&=&\vec{n}\cdot({\vec{x}}_T\times{\vec{p}}\ )\Delta F(x_T,E),
\end{eqnarray}
where $\vec{n}$ is the polarization vector for $q_1$ in its rest frame.
$F(x_T,E)$ and $\Delta F(x_T,E)$ are functions of both
$x_T\equiv|\vec{x}_T|$ and the c.m. energy $E$ of the two quarks.
We can show that the quark-quark scattering
with HTL propagators has the same form as that
in the static potential model \cite{Liang:2004ph}.
But the detailed expressions of $F(x_T,\hat{s})$ and
$\Delta F(x_T,\hat{s})$ are much more complicated.

In fact, one can show that these two parts of the cross sections
should have the same form as given in Eqs. (\ref{csgform}) and (\ref{dcsgform})
due to parity conservation in the scattering process.
We note that in an unpolarized reaction,
the cross section should be independent of any transverse direction.
Hence $d\sigma/d^2\vec x_T$ depends only on the magnitude of $x_T$
but not on the direction.
For the spin-dependent part, the only scalar that we can construct from
the available vectors is $\vec n\cdot(\vec p\times\vec x_T)$.
%This is why it has to take the form in Eqs. (\ref{csgform}-\ref{dcsgform}).

We note that, $\vec{x}_T\times\vec{p}$ is nothing but the
relative orbital angular momentum of the two-quark system,
$\vec{l}=\vec{x}_T\times\vec{p}$. Therefore, the polarized cross
section takes its maximum when
$\vec n$ is parallel or antiparallel to the relative
orbital angular momentum, depending on whether $\Delta F$ is
positive or negative. This corresponds to quark polarization in the
direction $\vec l$ or $-\vec l$.

As discussed in the last section, the average relative orbital angular
momentum $\vec l$ of two scattering quarks is in the opposite direction
of the reaction plane in non-central $A+A$ collisions.
Since a given direction of $\vec l$ corresponds to a given direction
of $\vec x_T$, there should be a preferred direction
of $\vec x_T$ at a given direction of the nucleus-nucleus
impact parameter $\vec b$. The distribution of $\vec x_T$ at
given $\vec b$ depends on the collective longitudinal momentum
distribution shown in the last section.
For simplicity, we consider a uniform distribution of
$\vec x_T$ in all possible directions in the upper half $xy$-plane
with $x>0$. In this case, we need to
integrate $d\Delta\sigma/d^{2}\vec x_T$ and $d\sigma/d^{2}\vec x_T$
over the half plane above $y$-axis to obtain the
average cross section at a given $\vec b$, i.e.,
\begin{eqnarray}
\label{csaverage}
\sigma &=&\int_{0}^{+\infty}dx\int_{-\infty}^{+\infty}dy\quad
\frac{d^2\sigma}{d^{2}{\vec{x}}_{T}}, \\
%\end{equation}
%\begin{equation}
\label{dcsaverage}
\Delta\sigma &=&
\int_{0}^{+\infty}dx\int_{-\infty}^{+\infty}dy\quad
\frac{d^2\Delta\sigma}{d^{2}{\vec{x}}_{T}} .
\end{eqnarray}
The polarization of the quark is then obtained as,
\begin{equation}
P_q=\frac{\Delta\sigma}{\sigma}.
\end{equation}

\subsection{Small angle approximation}

We only consider light quarks and
neglect their masses. Carrying out the traces in Eq.(\ref{trace}),
we can obtain the expression of the cross section
with HTL gluon propagators.
The results are much more complicated than those
as obtained in Ref.\cite{Liang:2004ph}
using a static potential model.
However, if we use small angle or small transverse momentum
transfer approximation, the results are still very simple.
In this case, with $q_z\sim 0$ and $q_T\equiv |\vec q_T|\ll p$,
we obtain the spin-independent (unpolarized) cross section ,
\begin{equation}
\frac{d^2\sigma}{d^{2}{\vec x}_{T}}=\frac{g^4c_{qq}}{8}
\int\frac{d^{2}{\vec{q}}_{T}}{(2\pi)^{2}}\frac{d^{2}{\vec{k}}_{T}}{(2\pi)^{2}}
e^{i({\vec{k}}_{T}-{\vec{q}}_{T})\cdot\vec{x}_T}
(\frac{1}{q_T^2+\mu_m^2}+\frac{1}{q_T^2+\mu_D^2})
(\frac{1}{k_T^2+\mu_m^2}+\frac{1}{k_T^2+\mu_D^2}),
\label{eq:cssap1}
\end{equation}
and the spin-dependent differential (polarized) cross section,
\begin{eqnarray}
\label{eq:dcssap1}
\frac{d^2\Delta{\sigma}}{d^{2}{\vec x}_{T}}
&=&-i\frac{g^4c_{qq}}{16}
\int\frac{d^{2}{\vec q}_{T}}{(2\pi)^{2}}\frac{d^{2}{\vec k}_{T}}{(2\pi)^{2}}
e^{i(\vec{k}_T-\vec{q}_T)\cdot\vec{x}_T}
\frac{(\vec{k}_T-\vec{q}_T)\cdot(\vec{p}\times\vec{n})}{p^2}\nonumber\\
&&
%\phantom{XXXXXXXXXX}
\times (\frac{1}{q_T^2+\mu_m^2}+\frac{1}{q_T^2+\mu_D^2})
(\frac{1}{k_T^2+\mu_m^2}+\frac{1}{k_T^2+\mu_D^2}).
\end{eqnarray}
We note that the polarized differential cross can be related to
the unpolarized one by,
\begin{equation}
\frac{d^2\Delta\sigma}{d^{2}{\vec x}_{T}}=
-\frac{1}{2{p^2}}(\vec{p}\times\vec{n})\cdot
\vec{\nabla}_T\frac{d^2\sigma}{d^{2}{\vec x}_{T}}.
\end{equation}
Completing the integration over the transverse momentum transfer,
\begin{equation}
\int\frac{d^2\vec q_T}{(2\pi)^2} \frac{e^{i\vec q_T\cdot \vec x_T}}{q_T^2+\mu_m^2}
=\int\frac{q_Tdq_T}{2\pi}\frac{J_0(q_Tx_T)}{q_T^2+\mu_m^2},
\label{intform1}
\end{equation}
\begin{equation}
\int_{0}^{\infty}q_Tdq_T\frac{J_0(q_Tx_T)}{q_T^2+\mu_m^2}=K_{0}(\mu_mx_T),
\label{intform2}
\end{equation}
where $J_0$ and $K_0$ are the Bessel and modified
Bessel functions respectively,
we obtain,
\begin{equation}
\frac{d^2\sigma}{d^{2}{\vec x}_{T}}=
%\frac{g^4c_{qq}}{8}
%\int\frac{dq_T}{2\pi}\frac{dk_T}{2\pi}
%q_T J_0(q_Tx_T)k_T J_0(k_Tx_T)
%&&\times
%(\frac{1}{q_T^2+\mu_m^2}+\frac{1}{q_T^2+\mu_D^2})
%(\frac{1}{k_T^2+\mu_m^2}+\frac{1}{k_T^2+\mu_D^2})\nonumber\\
\frac{c_{qq}}{2}\alpha_s^2
[K_0(\mu_{m}x_T)+K_0(\mu_{D}x_T)]^{2},
\end{equation}
\begin{equation}
\frac{d^2\Delta\sigma}{d^{2}{\vec x}_{T}}
%&=&-i\frac{g^4c_{qq}}{16}
%\int\frac{dq_T}{2\pi}\frac{dk_T}{2\pi}
%\frac{(\vec{k}_T-\vec{q}_T)\cdot(\vec{p}\times\vec{n})}{p^2}
%q_T J_0(q_Tx_T)k_T J_0(k_Tx_T)\nonumber\\
%&&\times
%(\frac{1}{q_T^2+\mu_m^2}+\frac{1}{q_T^2+\mu_D^2})
%(\frac{1}{k_T^2+\mu_m^2}+\frac{1}{k_T^2+\mu_D^2})\nonumber\\
=\frac{c_{qq}\alpha_s^2}{2}
\frac{(\vec{p}\times\vec{n})\cdot\hat{x}_T}{p^2}
[K_0(\mu_{m}x_T)+K_0(\mu_{D}x_T)]
[\mu_{m}K_1(\mu_{m}x_T)+\mu_{D}K_1(\mu_{D}x_T)],
\end{equation}
where $\hat{x}_T=\vec x_T/x_T$ is the unit vector of $\vec x_T$.
We compare the above results with that in the
screened static potential model (SPM) where one also made the
small angle approximation,
\begin{eqnarray}
\Bigl[\frac{d\sigma}{d^{2}{\vec x}_{T}}\Bigr]_{SPM}&=&\frac{g^4c_{T}}{4}
\int\frac{d^{2}{\vec{q}}_{T}}{(2\pi)^{2}}\frac{d^{2}{\vec{k}}_{T}}{(2\pi)^{2}}
e^{i({\vec{k}}_{T}-{\vec{q}}_{T})\cdot\vec{x}_T}
\frac{1}{q_T^{2}+\mu_D^2}\frac{1}{k_T^{2}+\mu_D^2},\\
\Bigl[\frac{d\Delta{\sigma}}{d^{2}{\vec x}_{T}}\Bigr]_{SPM}
&=&-i\frac{g^4c_T}{8} \int\frac{d^{2}{\vec q}_{T}}{(2\pi)^{2}}\frac{d^{2}{\vec k}_{T}}{(2\pi)^{2}}
e^{i(\vec{k}_T-\vec{q}_T)\cdot\vec{x}_T}
\frac{(\vec{k}_T-\vec{q}_T)\cdot(\vec{p}\times\vec{n})}
{p^2(q_T^{2}+\mu_D^2)(k_T^{2}+\mu_D^2)}.
\end{eqnarray}
We see that the only difference between the two results is
the additional contributions from magnetic gluons,
whose contributions are absent in the static potential model.
Using Eqs. (\ref{intform1}) and (\ref{intform2}),
we recover the results in Ref. \cite{Liang:2004ph},
\begin{equation}
\Bigl[\frac{d\sigma}{d^{2}{\vec x}_{T}}\Bigr]_{SPM}
=\alpha_s^2c_T K_0^2(\mu_{D}x_T),
\end{equation}
\begin{equation}
\Bigl[\frac{d\Delta\sigma}{d^{2}{\vec x}_{T}}\Bigr]_{SPM}
=\alpha_s^2c_T
\frac{(\vec{p}\times\vec{n})\cdot\hat{\vec{x}}_T}{p^2}
\mu_D K_0(\mu_{D}x_T)K_1(\mu_{D}x_T)).
\end{equation}

\subsection{Beyond small angle approximation}

Now we present the complete results for the cross-section in
impact parameter space using HTL gluon propagators without
small angle approximation. The unpolarized and polarized cross
section can be expressed in general as,
%slightly simple
%and can be given as in the following.
\begin{equation}
\frac{d\sigma}{d^{2}{\vec x}_{T}}= \frac{g^4c_{qq}}{64p^2}
\int\frac{d^{2}{\vec{q}}_{T}}{(2\pi)^{2}}\frac{d^{2}{\vec{k}}_{T}}{(2\pi)^{2}}
e^{i({\vec{k}}_{T}-{\vec{q}}_{T})\cdot\vec{x}_T}
\frac{f(\vec{q}_T,\vec{k}_T)}{\Lambda(\vec q_T)\Lambda(\vec k_T)},
\end{equation}
\begin{equation}
\frac{d\Delta{\sigma}}{d^{2}{\vec x}_{T}}= -i\frac{g^4c_{qq}}{64p^3}
\int\frac{d^{2}{\vec q}_{T}}{(2\pi)^{2}}\frac{d^{2}{\vec k}_{T}}{(2\pi)^{2}}
e^{i(\vec{k}_T-\vec{q}_T)\cdot\vec{x}_T}
\frac{\Delta f(\vec{q}_T,\vec{k}_T)}{\Lambda(\vec q_T)\Lambda(\vec k_T)},
\end{equation}
where the kinematic factor becomes
$\Lambda(\vec q_T)=\sqrt{2p|p_{3z}^+|}$;
$f$ and $\Delta f$ are given by,
\begin{eqnarray}
f({\vec q}_{T},{\vec k}_{T}) &=&
\frac{A_{mm}({\vec q}_{T},{\vec{k}}_{T})}{(q^2+\mu_m^2)(k^2+\mu_m^2)}
+\frac{A_{ee}({\vec q}_{T},{\vec{k}}_{T})}{(q^2+\mu_D^2)(k^2+\mu_D^2)}
\nonumber\\
& & +\frac{A_{me}({\vec q}_{T},{\vec{k}}_{T})}{(q^2+\mu_m^2)(k^2+\mu_D^2)}
+\frac{A_{me}({\vec k}_{T},{\vec{q}}_{T})}{(q^2+\mu_D^2)(k^2+\mu_m^2)};
\end{eqnarray}
\begin{eqnarray}
\Delta f({\vec q}_{T},{\vec k}_{T})&=&
\frac{\Delta A_{mm}({\vec{q}}_{T},{\vec{k}}_{T})}{(q^2+\mu_m^2)(k^2+\mu_m^2)}
+ \frac{\Delta A_{ee}({\vec q}_{T},{\vec{k}}_{T})}{(q^2+\mu_D^2)(k^2+\mu_D^2)}
\nonumber\\
& &
+\frac{\Delta A_{me}({\vec{q}}_{T},{\vec{k}}_{T})}{(q^2+\mu_m^2)(k^2+\mu_D^2)}
-\frac{\Delta A_{me}({\vec{k}}_{T},{\vec{q}}_{T})}{(q^2+\mu_D^2)(k^2+\mu_m^2)},
\end{eqnarray}
\begin{eqnarray}
A_{mm}({\vec q}_{T},{\vec k}_{T}) &=&
(\vec{q}\cdot\vec{k})^2+8p^2(\vec{q}\cdot\vec{k})+8p^3(q_z+k_z)+16p^4;  \\
%
%A_{ee}({\vec q}_{T},{\vec k}_{T}) &=&
%p^2[(\vec{q}\cdot\vec{k})^2+4p(q_z+k_z)(\vec{q}\cdot\vec{k})+8p^2(\vec{q}\cdot\vec{k})
%+4p^2(q_z+k_z)^2                                                           \nonumber\\
%& &+16p^3(q_z+k_z) +16p^4]                                                    \nonumber\\\
%
A_{ee}({\vec q}_{T},{\vec k}_{T}) &=&
A_{mm}({\vec q}_{T},{\vec k}_{T})+
4p(q_z+k_z)[(\vec{q}\cdot\vec{k})+p(q_z+k_z)+2p^2]; \\
A_{me}({\vec{q}}_{T},{\vec{k}}_{T}) &=&
A_{mm}({\vec q}_{T},{\vec k}_{T})+
[-2q_z k_z(\vec{q}\cdot\vec{k})+4pk_z(\vec{q}\cdot\vec{k})-2pq_z^2k_z \nonumber\\
& & -2pq_z{k_z}^2+4p^2{k_z}^2+8p^3 k_z)];  \\
%A_{Dm}({\vec q}_{T},{\vec k}_{T}) &=&
%A_{me}({\vec k}_{T},{\vec q}_{T})                       \nonumber\\
\Delta A_{mm}({\vec{q}}_{T},{\vec{k}}_{T}) &=&
-\{[\vec{q}\cdot\vec{k}+4p^2-2p(q_z+k_z)](k_z\vec{q}_T-q_z\vec{k}_T) \nonumber\\
& & +2p(\vec{q}\cdot\vec{k}+4p^2)(\vec{q}_T-\vec{k}_T)\}\cdot(\vec{p}\times\vec{n}); \\
\Delta A_{ee}({\vec{q}}_{T},{\vec{k}}_{T}) &=&
\Delta A_{mm}({\vec{q}}_{T},{\vec{k}}_{T})-
4p(q_z+k_z)[(k_z\vec{q}_T-q_z\vec{k}_T)-p(\vec{k}_T-\vec{q}_T)]
\cdot(\hat{p}\times\vec{n});\phantom{XX} \\
\Delta A_{me}({\vec{q}}_{T},{\vec{k}}_{T}) &=&
\Delta A_{mm}({\vec{q}}_{T},{\vec{k}}_{T})
+2pk_z[2p(\vec{k}_T-\vec{q}_T)+(q_z-k_z){\vec{q}}_T
\nonumber\\
&&-(k_z\vec{q}_T-q_z\vec{k}_T)]\cdot(\hat{p}\times\vec{n})
+2q_zk_z(k_z\vec{q}_T-q_z\vec{k}_T)\cdot(\hat{p}\times\vec{n}),
%\\
%\Delta{A}_{Dm}({\vec q}_{T},{\vec k}_{T}) &=&
%-\Delta{A}_{me}({\vec{k}}_{T},{\vec{q}}_{T})
%\nonumber\
\end{eqnarray}
where $p\equiv|\vec p|$.
%, $q\equiv|\vec q|$, and $k\equiv|\vec k|$.
It is useful to note that
\begin{equation}
A_{mm}(\vec q_T,\vec k_T)=A_{mm}(\vec k_T,\vec q_T),
\end{equation}
\begin{equation}
A_{ee}(\vec q_T,\vec k_T)=A_{ee}(\vec k_T,\vec q_T).
\end{equation}
Hence, $f(\vec q_T,\vec k_T)$ is symmetric in its two variables
\begin{equation}
f(\vec q_T,\vec k_T)=f(\vec k_T,\vec q_T).
\end{equation}
Similarly from
\begin{equation}
\Delta A_{mm}(\vec q_T,\vec k_T)=-\Delta A_{mm}(\vec k_T,\vec q_T),
\end{equation}
\begin{equation}
\Delta A_{ee}(\vec q_T,\vec k_T)=-\Delta A_{ee}(\vec k_T,\vec q_T),
\end{equation}
we know that $\Delta f(\vec q_T,\vec k_T)$ is anti-symmetric,
\begin{equation}
\Delta f(\vec q_T,\vec k_T)=-\Delta f(\vec k_T,\vec q_T).
\end{equation}

As mentioned above, to get the average polarization for a fixed
direction of the reaction plane in heavy-ion collisions,
we need to average over the distribution of $\vec x_T$.
For this purpose, we take the approach as in Ref. \cite{Liang:2004ph},
and integrate $d\sigma/d^{2}\vec x_T$ and $d\Delta\sigma/d^{2}\vec x_T$
over the half plane above the $y$-axis as shown in Eqs. (\ref{csaverage})
and (\ref{dcsaverage}).
It is convenient to carry out first
the integration over $x$ and $y$ then that over $\vec q_T$ and $\vec k_T$.
To do this, we use the identity,
\begin{equation}
2\int_{0}^{+\infty}\frac{dx}{2\pi}\int_{-\infty}^{+\infty}\frac{dy}{2\pi}
 e^{i({\vec{k}}_{T}-{\vec{q}}_{T})\cdot\vec{x}_T}=
\delta^2(\vec k_T-\vec q_T)
+\frac{i}{\pi}\delta(k_y-q_y)\mathcal{P}\frac{1}{k_x-q_x},
\label{intiden}
\end{equation}
where $\mathcal{P}$ denotes the principal value.

It is useful to note that
$\delta^2(\vec k_T-\vec q_T)=\delta^2(\vec q_T-\vec k_T)$
and $\mathcal{P}\frac{1}{k_x-q_x}
=-\mathcal{P}\frac{1}{q_x-k_x}$.
Therefore, only the first term on the r.h.s. of Eq. (\ref{intiden})
contributes to the total unpolarized cross section,
\begin{equation}
\sigma=\frac{g^4 c_{qq}}{64p^2}\frac{1}{2} \int_{q_T\le
p}\frac{d^{2}{\vec{q}}_{T}}{(2\pi)^{2}}
\frac{f(\vec{q}_T,\vec{q}_T)}{\Lambda^2(\vec q_T)}.
\end{equation}
%This is nothing but the total cross section in the unpolarized case.
The polarized cross section $\Delta\sigma$ receives contribution only
from the second term,
\begin{equation}
\Delta{\sigma}=\frac{g^4c_{qq}}{64p^3}
\int_{-p}^{p}\frac{dq_y}{2\pi}
\int_{-\sqrt{p^2-{q_y}^2}}^{\sqrt{p^2-{q_y}^2}}\frac{dq_x}{2\pi}
\int_{-\sqrt{p^2-{q_y}^2}}^{\sqrt{p^2-{q_y}^2}}\frac{dk_x}{2\pi}
\frac{1}{k_x-q_x}\frac{\Delta f(q_x,q_y;k_x,q_y)}{\Lambda(\vec
q_T)\Lambda(\vec k_T)}.
\end{equation}
Changing the integration variable $q_T=p\sin\theta$
and $\xi=\sin^2(\theta/2)$ in the expression of the total
cross section $\sigma$, we obtain,
\begin{equation}
\sigma = \frac{\pi c_{qq}\alpha_s^2}{4\hat s} \int_{0}^{1}d\xi
\left\{\frac{1+\xi^2}{(\xi+\beta_m\tilde{T}^2)^2}
+\frac{(1-\xi)^2}{(\xi+\beta_D\tilde{T}^2)^2} +\frac{2(1-\xi)}
{(\xi+\beta_D\tilde{T}^2)(\xi+\beta_m\tilde{T}^2)} \right\},
\end{equation}
where $\beta_D=(\mu_D/T)^2=4\pi\alpha_s(N_c+N_f/2)/3$,
$\beta_m=(\mu_m/T)^2=0.255^2(4\pi)^2\alpha_s^2N_c/2$,
and $\tilde{T}=T/\sqrt{\hat{s}}$ with
$\sqrt{\hat{s}}$ the center of mass energy of the $qq$-system.
The integration can be carried out analytically,
\begin{eqnarray}
\label{csres}
\sigma&=&\frac{\pi c_{qq}\alpha_s^2}{4\hat s}
\left\{4+\frac{1}{\beta_m\tilde{T}^2}+\frac{1}{\beta_D\tilde{T}^2}
-\frac{2}{1+\beta_m\tilde{T}^2}
-2(\beta_m\tilde{T}^2+
\frac{1+\beta_m\tilde{T}^2}{\beta_m\tilde{T}^2-\beta_D\tilde{T}^2})
\ln(1+\frac{1}{\beta_m\tilde{T}^2})\right.\nonumber\\
& &\left.-2(\beta_D\tilde{T}^2+
\frac{\beta_m\tilde{T}^2-2\beta_D\tilde{T}^2-1}{\beta_m\tilde{T}^2-\beta_D\tilde{T}^2})
\ln(1+\frac{1}{\beta_D\tilde{T}^2})\right\}.
\end{eqnarray}
Similarly, we make the variable substitutions
$q_y=p\sqrt{1-t^2}$, $q_x=pt\sqrt{1-\xi^2}$, $k_x=pt\sqrt{1-\eta^2}$ in
the integration for $\Delta{\sigma}$ and obtain,
\begin{eqnarray}
\label{dcsres}
\Delta\sigma
&&= -\frac{c_{qq}\alpha_s^2}{8\pi\hat s}
\int_{-1}^{1}dt\int_{0}^{1}d\xi\int_{0}^{1}d\eta
\frac{t^2\sqrt{\xi\eta}}{\sqrt{1-t^2}\sqrt{1-\xi^2}\sqrt{1-\eta^2}}  \nonumber\\
&& \times\left\{
\frac{(1-t^2)(4+t\xi+t\eta)-2(t^2\xi\eta+1)+2t(1+\xi\eta)(5+t^2\xi\eta)/(\xi+\eta)}
{(1-t\xi+2\beta_m\tilde{T}^2)(1-t\eta+2\beta_m\tilde{T}^2)}\right.    \nonumber\\
&&\ \ \ \left.
+\frac{(1-t^2)(t\xi+t\eta)+2(t^2\xi\eta+1)+2t(1+\xi\eta)(1+t^2\xi\eta)/(\xi+\eta)}
{(1-t\xi+2\beta_D\tilde{T}^2)(1-t\eta+2\beta_D\tilde{T}^2)}\right.    \nonumber\\
&&\ \ \ \left.
+\frac{2(1-t^2)(2+t\xi+t\eta)+8t(1+\xi\eta)(1+t\eta)/(\xi+\eta)}
{(1-t\xi+2\beta_m\tilde{T}^2)(1-t\eta+2\beta_D\tilde{T}^2)} \right\}.
\end{eqnarray}
Note that in the calculation of both the polarized and
unpolarized cross sections we have limited the range of integration
over the transverse momentum due to energy conservation. Such restriction
is not imposed in the small angle approximation in
Ref.~\cite{Liang:2004ph}. We see that, at given $\beta_m$ and $\beta_D$,
both $\sigma$ and $\Delta\sigma$ are functions
of the variable $\tilde{T}=T/\sqrt{\hat{s}}$.
Since $\beta_m$ and $\beta_D$ depend on $\alpha_s$,
the polarization $P_q=\Delta\sigma/\sigma$ also depends
on the value of $\alpha_s$.

We can now carry out the integration numerically to get
the quark polarization $P_q$. Before we show the numerical results,
it is useful to look at two limits.

(1) {\em High energy limit}. At very high energies,
$\sqrt{\hat{s}}\gg T$ or $\tilde{T}\ll 1$, we have,
\begin{equation}
\sigma = \frac{\pi c_{qq}\alpha_s^2}{4\hat s\tilde{T}^2}
\left\{\frac{1}{(1+4\beta_m\tilde{T}^2)\beta_m}
+\frac{1}{(1+4\beta_D\tilde{T}^2)\beta_D}
+\frac{2}{\beta_D-\beta_m}
\ln\frac{\beta_D(1+4\beta_m\tilde{T}^2)}{\beta_m(1+4\beta_D\tilde{T}^2)}\right\},
\label{cslim1}
\end{equation}
{\small
\begin{equation}
\Delta\sigma
= -\frac{4c_{qq}\alpha_s^2}{\pi \hat s}
\int_{0}^{1}dt
\left[\frac{1}{\sqrt{t^2+4\beta_m\tilde{T}^2}}
\tan^{-1}{\sqrt{\frac{1-t^2}{t^2+4\beta_m\tilde{T}^2}}}
+\frac{1}{\sqrt{t^2+4\beta_D\tilde{T}^2}}
\tan^{-1}{\sqrt{\frac{1-t^2}{t^2+4\beta_D\tilde{T}^2}}}
\right]^2.
\label{dcslim1}
\end{equation}}
This is the case where the small angle approximation can be made.
The above can also be obtained from
Eqs. (\ref{eq:cssap1}-\ref{eq:dcssap1}) given in the last section
by carrying out the integration over $\vec x_T$ in the half plane of $x>0$.

(2) {\em Low energy limit}.
In the limit $\sqrt{\hat{s}}\ll T$, we have
$q_T^2+\mu_D^2\approx \mu_D^2$ and $q_T^2+\mu_m^2\approx \mu_m^2$,
the cross sections become
\begin{eqnarray}
\label{cslim2}
\sigma
&=& \frac{c_{qq}\alpha_s^2}{8\hat s}\pi(\frac{\sqrt{\hat{s}}}{T})^4
\left(\frac{8}{3\beta_m^2}+\frac{2}{3\beta_D^2}+\frac{2}{\beta_m\beta_D}\right), \\
\Delta\sigma
&=& -\frac{c_{qq}\alpha_s^2}{16\pi\hat s}(\frac{\sqrt{\hat{s}}}{T})^4
\left[-\frac{1}{\beta_m^2}\left(\frac{1}{192}\Gamma^4(\frac{1}{4})+\frac{1}{2}\Gamma^4(\frac{3}{4})\right)
+\frac{1}{\beta_D^2}\left(\frac{1}{192}\Gamma^4(\frac{1}{4})+\Gamma^4(\frac{3}{4})\right)\right.\nonumber\\
& &\left.
%\phantom{XXXXXXX}
+\frac{2}{\beta_m\beta_D}\left(\frac{1}{192}\Gamma^4(\frac{1}{4})+\Gamma^4(\frac{3}{4})\right)
\right].
\end{eqnarray}
Given the corresponding values of the $\Gamma$-function, one
can obtain numerically,
\begin{equation}
\Delta\sigma\approx-\frac{c_{qq}\alpha_s^2}{16\pi\hat{s}}(\frac{\sqrt{\hat{s}}}{T})^4
\left(-\frac{2.03}{\beta_m^2}+\frac{3.15}{\beta_D^2}+\frac{6.30}{\beta_m\beta_D}\right).
\label{dcslim2}
\end{equation}
We see that in the low energy limit the magnetic part contributes
with different sign from the electric one.
The polarization $P_q=\Delta\sigma/\sigma$ is given by
\begin{equation}
P_q\approx-\frac{3}{2\pi^2}
\frac{-2.03\beta_D^2+3.15\beta_m^2+6.30\beta_m\beta_D}
     {8\beta_D^2+2\beta_m^2+6\beta_m\beta_D},
\label{pollim2}
\end{equation}
which tends to be a constant in this low energy limit. In the weak
coupling limit $\alpha_s\ll 1$, $\beta_D\gg \beta_m$ the above constant
$P_q\approx 0.04$ is a small positive number.

%%% I don't touch this part --Wang Qun

It is also interesting to look at the contributions from the electric
part only. The corresponding cross sections, denoted with subscription $E$,
are
\begin{equation}
\left(\frac{d\sigma}{d^{2}{\vec x}_{T}}\right)_E
= \frac{g^4c_{qq}}{64p^2}
\int\frac{d^{2}{\vec{q}}_{T}}{(2\pi)^{2}}\frac{d^{2}{\vec{k}}_{T}}{(2\pi)^{2}}
\frac{A_{ee}({\vec q}_{T},{\vec{k}}_{T})e^{i({\vec{k}}_{T}-{\vec{q}}_{T})\cdot\vec{x}_T}}
{\Lambda(\vec q_T)\Lambda(\vec k_T)(q^2+\mu_D^2)(k^2+\mu_D^2)},
\end{equation}
\begin{equation}
\left(\frac{d\Delta{\sigma}}{d^{2}{\vec x}_{T}}\right)_E
= -i\frac{g^4c_{qq}}{64p^3}
\int\frac{d^{2}{\vec q}_{T}}{(2\pi)^{2}}\frac{d^{2}{\vec k}_{T}}{(2\pi)^{2}}
\frac{\Delta A_{ee}({\vec q}_{T},{\vec{k}}_{T})e^{i(\vec{k}_T-\vec{q}_T)\cdot\vec{x}_T}}
{\Lambda(\vec q_T)\Lambda(\vec k_T)(q^2+\mu_D^2)(k^2+\mu_D^2)}.
\end{equation}
Carrying out the integration over $d^2\vec x_T$ in the
half plane with $x>0$, we obtain,
\begin{equation}
\label{csres_e}
\sigma_E =\frac{\pi c_{qq}\alpha_s^2}{4\hat s}\int_{0}^{1}
\frac{(1-\xi)^2d\xi}{(\xi+\beta_D\tilde{T}^2)^2}
       =\frac{\pi c_{qq}\alpha_s^2}{4\hat s}
\left[2+\frac{1}{\beta_D\tilde{T}^2}
-2(1+\beta_D\tilde{T}^2)\ln(1+\frac{1}{\beta_D\tilde{T}^2})\right],
\end{equation}
\begin{eqnarray}
\label{dcsres_e}
\Delta\sigma_E
&&= -\frac{c_{qq}\alpha_s^2}{8\pi\hat s}
\int_{-1}^{1}dt\int_{0}^{1}d\xi\int_{0}^{1}d\eta
\frac{t^2\sqrt{\xi\eta}}{\sqrt{1-t^2}\sqrt{1-\xi^2}\sqrt{1-\eta^2}}  \nonumber\\
&& \times\left\{
\frac{(1-t^2)(t\xi+t\eta)+2(t^2\xi\eta+1)+2t(1+\xi\eta)(1+t^2\xi\eta)/(\xi+\eta)}
{(1-t\xi+2\beta_D\tilde{T}^2)(1-t\eta+2\beta_D\tilde{T}^2)} \right\}.
\end{eqnarray}

In the high energy limit, where small angle approximation is
applicable, we have,
\begin{eqnarray}
\sigma_E
&=& \frac{\pi c_{qq}\alpha_s^2}
{4\hat s\beta_D\tilde{T}^2(1+4\beta_D\tilde{T}^2)},
\label{cslim1_e}
\\
\Delta\sigma_E
&=& -\frac{4c_{qq}\alpha_s^2}{\pi \hat s}
\int_{0}^{1}\frac{dt}{t^2+4\beta_D\tilde{T}^2}
\left[\tan^{-1}{\sqrt{\frac{1-t^2}{t^2+4\beta_D\tilde{T}^2}}}\ \right]^2.
\label{dcslim1_e}
\end{eqnarray}
In the low energy limit, we have,
\begin{eqnarray}
\label{cslim2_e}
\sigma_E
&=& \frac{c_{qq}\alpha_s^2}{12\hat s\beta_D^2}\pi(\frac{\sqrt{\hat{s}}}{T})^4, \\
\Delta\sigma_E
&=& -\frac{c_{qq}\alpha_s^2}{16\pi\hat s}(\frac{\sqrt{\hat{s}}}{T})^4
\frac{1}{\beta_D^2}
\left(\frac{1}{192}\Gamma^4(\frac{1}{4})+\Gamma^4(\frac{3}{4})\right).
\end{eqnarray}
The polarization in this case,
\begin{equation}
P_q^E \equiv {\Delta\sigma_E}/{\sigma_E}
=-\frac{3}{4\pi^2}
\left(\frac{1}{192}\Gamma^4(\frac{1}{4})+\Gamma^4(\frac{3}{4})\right)
\approx -0.24
\end{equation}
is a negative constant which can also be obtained
from Eq.~(\ref{pollim2}) by taking the limit $\beta_m\gg \beta_D$.

\subsection{Numerical results}

\begin{figure}[htbp]
\includegraphics[width=12cm]{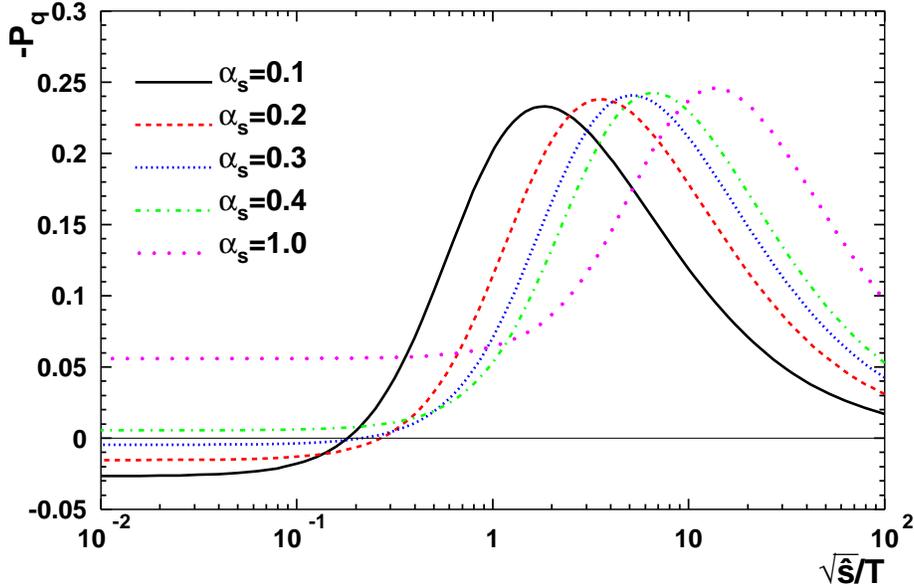}
%\centerline{\psfig{figure=geo2.eps,width=3.0in,height=2.4in}}
\caption{(Color online) Quark polarization -$P_q$ as a function of
$\sqrt{\hat{s}}/T$ for different $\alpha_s$'s. }
\label{pol1}
\end{figure}

We now carry out the integration in Eq. (\ref{dcsres}) numerically
and obtain the results for the quark polarization at intermediate
energies between the high-energy and low-energy limit.
The results are shown in Fig.~\ref{pol1} as functions of $\sqrt{\hat{s}}/T$.
The quark polarization ($-P_q$) along the reaction
plane approaches a small negative value as we have shown in the last
subsection in the low-energy limit. The value of the low energy limit
varies with $\alpha_s$ as given by Eq. (\ref{pollim2}). Such a
dependence on $\alpha_s$ is a consequence of the magnetic and electric
screening masses in the polarized and unpolarized cross sections
which have different dependence on $\alpha_s$. However, from
Eq. (\ref{pollim2}), the low-energy limit of the quark polarization
becomes independent of $\alpha_s$ in the weak coupling
limit $\alpha_s\rightarrow 0$ when $\beta_m\ll \beta_D$.

\begin{figure}[htbp]
\includegraphics[width=12cm]{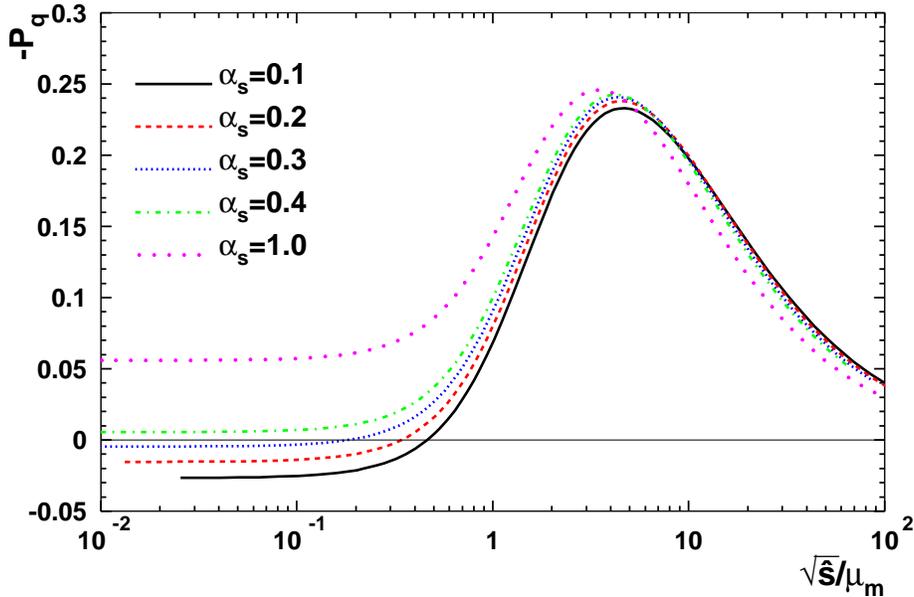}
%\centerline{\psfig{figure=geo2.eps,width=3.0in,height=2.4in}}
\caption{(Color online) Quark polarization -$P_q$ as a function of
$\sqrt{\hat{s}}/\mu_m$ for different values of $\alpha_s$. }
\label{pol2}
\end{figure}

As one increases the relative c.m. energy, the quark
polarization changes drastically with $\sqrt{\hat{s}}/T$.
It increases to some maximum values and then
decreases with the growing energy, approaching
the result of small angle approximation in the high-energy limit.
This structure is caused by the interpolation
between the high-energy and low-energy
behavior dominated by the magnetic part of the interaction
in the weak coupling limit $\alpha_s<1$. Therefore, the position
of the maxima in $\sqrt{\hat s}$ should
approximately scale with the magnetic mass $\mu_m$. This is
indeed the case as shown in Fig. \ref{pol2}.

\begin{figure}[htbp]
\includegraphics[width=12cm]{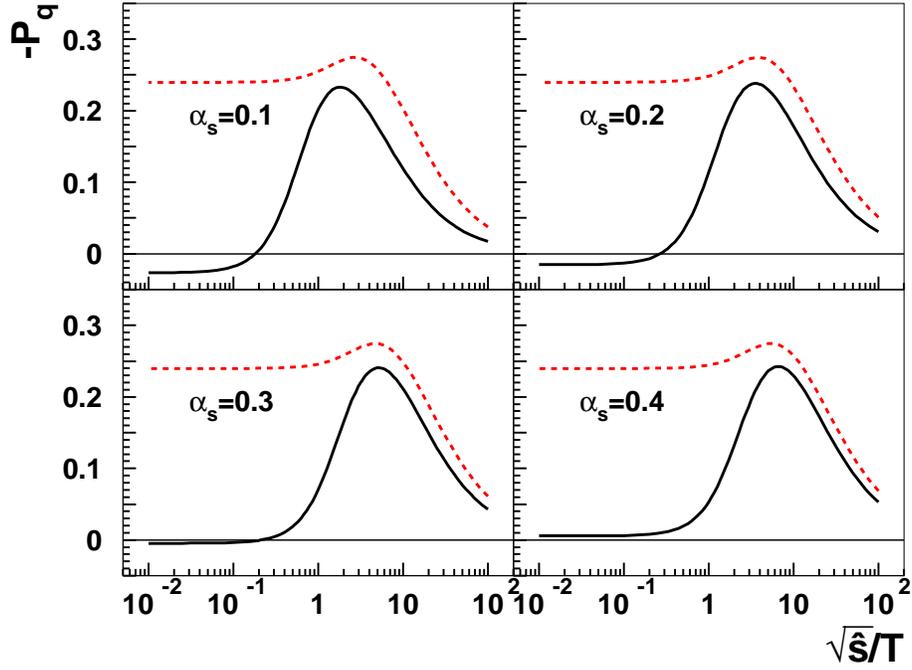}
%\centerline{\psfig{figure=geo2.eps,width=3.0in,height=1.6in}}
\caption{(Color online) Quark polarization -$P_q$ as a function of
$\sqrt{\hat{s}}/T$ with the full HTL gluon propagator (solid)
as compared to the results with the electric part of
the interaction only (dashed).}
\label{pol4}
\end{figure}

To further understand the interpolation between the high and low-energy
limits in the numerical results, we also compare them
in Fig. \ref{pol4} with the results with the electric gluon exchange only.
Without the contribution from the magnetic gluon interaction, the quark
polarization takes a relatively large value $P_q\approx -0.24$
at low energies and then decreases with $\sqrt{\hat s}$
at high energies. The magnetic interaction in the low-energy
limit apparently has a different sign in the contribution to the polarized
cross section relative to that of the electric one.
The net polarization is therefore reduced at finite
$\alpha_s$ to smaller negative values when
$\alpha_s\ll 1$. The electric contribution to the
net quark polarization also corresponds to the limit $\mu_m\gg\mu_D$
or $\alpha_s\gg 1$ in the full result. Even though our perturbative
approach is no longer valid in such a limit, it indicates
that the net quark polarization remains a finite negative value in
the strong coupling limit as shown in Fig. \ref{pol2}.

\begin{figure}[htbp]
\includegraphics[width=12cm]{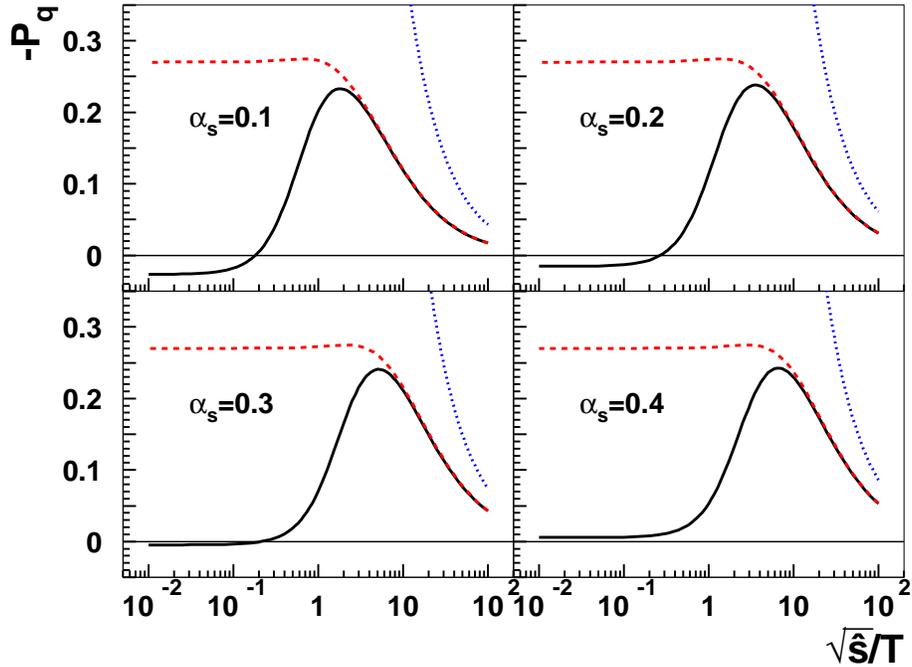}
%\centerline{\psfig{figure=geo2.eps,width=3.0in,height=2.4in}}
\caption{(Color online) Comparison of the results obtained using HTL gluon propagator
(solid line) with those under small angle approximation (dashed line)
and those using the screened static potential model under small angle approximation
(dotted line). }
\label{pol3}
\end{figure}

In Fig. \ref{pol3} we also compare the full numerical
results (solid lines) with those of the small angle
approximation in the high-energy limit (dashed lines)
as given by Eqs. (\ref{cslim1}) and (\ref{dcslim1}).
These two groups of results indeed agree with each other
at high energies. However, they both are different from
the results of the static potential
model in the small angle limit (dotted lines) \cite{Liang:2004xn}
which does not have the energy conservation restriction in the
integration over the transverse momentum transfer.

\section{Conclusions and discussions}

In this paper, we have extended an earlier study \cite{Liang:2004xn} of
the global quark polarization caused by the longitudinal fluid shear
in non-central heavy-ion collisions. We have calculated the
average local relative orbital angular momentum or longitudinal
fluid shear with two extreme models: Landau fireball and Bjorken scaling
model. In the Landau fireball model, we assumed a wounded nucleon
model for local particle production with both the hard-sphere
and Woods-Saxon nuclear distributions. Each parton is then assumed to
carry an average longitudinal flow velocity calculated from the
net longitudinal momentum at a given transverse position.
In the Bjorken scaling model we considered correlation between
spatial and momentum rapidity in a 3-dimensional expanding system
for which we calculated the average rapidity or longitudinal momentum
shear (derivative of the average rapidity or the longitudinal momentum)
with respect to the transverse position $x$. The shear
determines the local relative orbital angular momentum in the
comoving frame at a given rapidity. These two model calculations
provide estimates of the local fluid shear in two extreme limits.

We have also extended the calculation of the global quark
polarization $P_q$ within perturbative QCD at finite temperature
beyond the small angle approximation of the previous
study \cite{Liang:2004xn} which might not be valid for
small values of the local longitudinal fluid shear or the
average c.m. energy $\sqrt{\hat s}$ of a colliding quark pair.
We found that the magnetic part of the interaction in one-gluon
exchange is particularly important at low energies
which cancels the contribution from
the electric interaction and leads to smaller negative
values of the net quark polarization in the weak coupling
limit ($\alpha_s<1$). The final global quark polarization
therefore is small in both the low and high-energy limits.
It can, however, reach a peak value of about $P_q\approx -0.24$
at an energy determined by the nonperturbative
magnetic mass $\sqrt{\hat s}\sim 4 \mu_m\approx  g^2 T \sqrt{N_c/2}$.
For $\sqrt{\hat s}< \mu_m$, the average quark polarization
becomes significantly smaller.

In semi-peripheral $Au+Au$ collisions ($b=R_A$) at the RHIC energy, one
can assume an average temperature $T\approx 400 $ MeV \cite{Muller:2005en}.
With $\alpha_s\approx 0.3$, the global quark polarization reaches its
peak value at c.m. energy about 1.8 GeV. Since the magnetic
interaction dominates the quark-quark interaction in our calculation,
we can assume that the average interaction range in the transverse direction
is given by the magnetic mass, $\Delta x\sim 1/\mu_m$. According to
our estimates of the longitudinal fluid shear, the average
c.m. energy of the quark pair under such fluid shear is
$\sqrt{\hat s}\sim 0.8~{\rm GeV}^2/\mu_m \approx 0.17$ GeV
(from Fig. \ref{fig:ly}) in the Landau fireball
model. In the Bjorken scaling model (from Fig. \ref{dndydx}), the
c.m. energy provided by the local fluid shear
is $\sqrt{\hat s}\sim 0.004\langle p_T\rangle \cosh (y)~{\rm fm}^{-1}/\mu_m
\approx 0.001$ GeV in the central rapidity region
(we assume $\langle p_T\rangle \sim 2T$). In both cases,
the longitudinal fluid shear is so weak that
the global quark polarization due to perturbative quark-quark
scatterings is quite small according to our numerical calculations
that go beyond the small angle approximation.

In heavy-ion collisions at the Large Hadron Collider (LHC) energy
$\sqrt{s}=5.5$ TeV, the average multiplicity density per participant
nucleon pair was estimated to be about a factor of 3 larger
than that at the RHIC energy \cite{Li:2001xa}. The corresponding
longitudinal fluid shear and the average c.m. energy of a quark pair
will be about a factor 6 larger than that at the RHIC energy in the
Landau fireball model, assuming the temperature is about 1.44
higher. One can also expect the average local longitudinal fluid
shear in the Bjorken scenario at LHC is
similarly amplified compared to the RHIC energy
in particular at large rapidity. Therefore, the resulting net quark
polarization should also be larger at LHC.

We want to emphasize that the above numerical estimate is based
on a perturbative calculation via quark-quark scatterings
in the weak coupling limit. It is
still possible that quarks could acquire large global
polarization through interaction in the strong coupling limit,
as hinted by our results with large values of the strong
coupling constant even though such a perturbative approach
becomes invalid. The finite value of the quark polarization
could be detected via measurements of the global hyperon polarization
or the vector meson spin alignment with respect to the reaction plane.
According to our estimate of the longitudinal fluid shear, the effect
is more significant at large rapidity under the
Bjorken scenario of the initial parton production.

In the limit of vanishing local orbital angular momentum provided by
the longitudinal fluid shear, the approach we used in this paper in the
impact-parameter representation might not be valid anymore. However,
the final spin-polarization due to the spin-orbital interaction should
approach to zero in this limit, which is approximately the result
of our full calculation.

\section{Acknowledgment}

The authors thank B. Muller for pointing out the scenario of
Bjorken scaling model of initial parton production and J.~Deng
for his help in the numerical calculation of the local orbital
angular momentum with the Woods-Saxon geometry.
This work was supported in part by
National Natural Science Foundation of China
(NSFC) under No. 10525523;
the startup grant
from University of Science and Technology of China (USTC)
in association with \emph{100 talents}
project of Chinese Academy of Sciences (CAS) and by NSFC
grant No. 10675109 and, the Director, Office of Energy Research,
Office of High Energy and Nuclear Physics, Divisions of
Nuclear Physics, of the U.S. Department of Energy
under No. DE-AC03-76SF00098.

%\end{multicols}

%\end{widetext}

\end{document}